\documentclass[a4paper,12pt]{article}

\usepackage{amsmath}
\usepackage{amssymb}
\usepackage{mathptmx}
\usepackage{amsfonts}
\usepackage[mathscr]{eucal}

\newcommand{\one}{{ \mathrm{1} \hspace{-0.25em}  \mathrm{l}}}

\newcommand{\mbf}[1]{\ensuremath{\mathbf{#1}}}
\newcommand{\mbb}[1]{\ensuremath{\mathbb{#1}}}

\newcommand{\Rthree}{\mathbb{R}^{\, 3} }
\newcommand{\lang}{\langle}
\newcommand{\rang}{\rangle}

\newcommand{\bos}{\mathrm{b}}

\newcommand{\tens}{\otimes}

\newcommand{\Ltwo}{L^{2}}
\newcommand{\ms}[1]{\ensuremath{\mathscr{#1}}}

\newcommand{\Fb}{\mathscr{F}_{\, \rm{b}}}

\newcommand{\sqz}{d  \Gamma }
\newcommand{\sqzb}{d  \Gamma_{\mathrm{b}} }
\newcommand{\Hp}{H_{\mathrm{p}}}
\newcommand{\Hb}{H_{\mathrm{b}}}
\newcommand{\HI}{H_{\mathrm{I}}}

\newcommand{\uR}{u_{{}_{R}}}
\newcommand{\chiR}{\chi_{{}_{R}}}

\newcommand{\aI}{a_{\mathrm{I}} }
\newcommand{\bI}{b_{ \mathrm{I} }}

\newtheorem{theorem}{Theorem}[section]
\newtheorem{proposition}[theorem]{Proposition}
\newtheorem{lemma}[theorem]{Lemma}
\newtheorem{corollary}[theorem]{Corollary}
\newtheorem{remark}{Remark}[section]

\begin{document}
\begin{center}
{\LARGE A remark on the binding condition by the decay of  particle's potential
 }\\
 $\;$ $\frac{}{} $ \\
 {\large  Toshimitsu Takaesu }  \\
 {\scriptsize $\;$ } \\
\textit{Cooperative Faculty of Education, Gunma University, Gunma, Japan}
\end{center}

\begin{quote}
\textbf{[Abstract]} 
The system of a  particle  interacting with a Bose field is investigated.
 It is proven  that  the binding condition holds by  the decay of particle's  potential. 
 As an application, the exponential decay of the ground state follows.
\end{quote}
\textit{Key words:} Quantum fields, Fock spaces, Ground states \\
\textit{2020 Mathematics Subject Classification Numbers:} 81Q10, 47B25

\section{Introduction and Main Result}
We consider the  system of  a non-relativistic  particle coupled to a scalar Bose field. 
The state space for the system is defined   by  
\begin{equation}
\ms{H}= \Ltwo ( \Rthree_{\mbf{x}} ) \tens \Fb (\Ltwo ( \Rthree_{\mbf{k}}  ) ) ,
\end{equation}
where $\Fb (\ms{K}) $ denotes the boson Fock space over  a Hilbert space $\ms{K} $. 
The total Hamiltonian of the system is defined by
\begin{equation}
H(\kappa ) = \Hp \tens \one + \one \tens \Hb + \kappa \HI .
\end{equation}
Here $\Hp = - \frac{\Delta }{2M}  +V $ and      $\Hb = \sqz (\omega )  $ where $\sqz (X )$ is the second quantization of $X$ and $\omega (\mbf{k}) = \sqrt{\mbf{k}^2 + m^2 } $, $m \geq 0 $.
The interaction is defined by $\HI = \frac{1}{\sqrt{2}} (a (\rho_{\mbf{x}}  ) + a^{\dagger} (\rho_{\mbf{x}} )  )$ where $a (f) $, $f \in L^2 (\Rthree )$,  is  the annihilation operator, and   $a^{\dagger }(g) $, $g \in L^2 (\Rthree )$, the creation operator. 
The function $\rho $ is defined  by
$\rho (\mbf{k}) = \frac{\mathbf{1}_{\Lambda} (| \mbf{k} | )}{\sqrt{\omega (\mbf{k})}}$, $\Lambda >0 $, 
where  $\mathbf{1}_{\Lambda} $ is  the definite function  on $ [ 0,  \Lambda ] \subset \mbb{R}$, and  we set $h_{\mbf{x}} (\mbf{k})
=  h(\mbf{k}) e^{-i \, \mbf{k} \cdot \mbf{x} }$ for  $h \in L^2 (\Rthree )$.
The creation operator and annihilation operator satisfy the canonical commutation relations:
\begin{align}
&[a(f) , a^{\dagger}(g)] = \lang f , g \rang , \\
&[a(f) , a (g)] = [a^{\dagger}(f) , a^{\dagger}(g) ] = 0 .
\end{align}
Let  $\Hp^0  = - \frac{1}{2M} \Delta $. Assume the condition below.
\begin{quote}
(A.1) $V=V(x)$ is  a real-valued function. There exist $0 \leq  \aI < 1$ and $0 \leq \bI $ such that for all $\psi \in \ms{D} (\Hp^0 ) $,   
\[
\| V \psi \| \leq  \aI \| \Hp^0  \psi \| + \bI \| \psi \|  . 
\]
\end{quote}
From (A.1), the Kato-Rellich theorem (\cite[Theorem X.12]{RS2}) yields that 
$\Hp $ is self-adjoint on $\ms{D} (H_0 )$ and bounded from below, in particular, $ \inf  \sigma  ( \Hp ) > - \max \left\{  \frac{\bI}{1-\aI} , \bI  \right\}$ holds. Here  $\sigma (X)$ denotes  the spectrum of operator $X$.

\begin{remark} 
We can also  define the particle Hamiltonian by  form. In this case,  we assume that 
there exist $0 \leq  a < 2M $ and $0 \leq b $ such that for all $\psi \in \cap_{j=1}^3 
\ms{D} ( \partial_{x_{j}} ) $,   
\[
 | \lang \psi , V \psi \rang | \leq  \frac{ a}{2M }  \lang   \nabla  \psi ,  \nabla  \psi \rang + b \lang  \psi , \psi \rang  .  \]
 Then, by the KLMN theorem (\cite[Theorem X.17]{RS2}),
 there exists a unique self-adjoint operator $\tilde{H}_{\mathrm{p}}$ such that 
 $\ms{D} (\tilde{H}_{\mathrm{p}}^{1/2})  = \cap_{j=1}^3 
\ms{D} ( \partial_{x_{j}} ) $ and for all $ \phi ,  \psi \in \ms{D} (\tilde{H}_{\mathrm{p}}^{1/2})$, 
\[
\lang  \tilde{H}_{\mathrm{p} }^{1/2} \phi , \tilde{H}_{\mathrm{p} }^{1/2} \psi \rang
= \frac{ 1 }{2M }  \lang   \nabla  \phi ,  \nabla  \psi \rang + \lang \phi , V \psi \rang . 
\]
\end{remark}
Let 
\[
H_0 =  \Hp \tens \one + \one \tens \Hb .
\]
Since the interaction $\HI $ is relatively bounded  to $\one \tens \Hb^{1/2}$, $\HI $ is relatively bounded to $H_0$ with infinitely small bound.
Hence,   $H(\kappa )$ is self-adjoint on $\ms{D} (H_0 )$ and bounded from below by the Kato-Rellich theorem (see, e.g., \cite[Theorem14.3]{Arai18}).
We define the total Hamiltonian without the particle's potential by
\begin{equation}
H^0 (\kappa ) = \Hp^0 \tens \one + \one \tens \Hb + \kappa \HI .
\end{equation}
Let $ \mbf{p} = - i \, \nabla $ and $\mbf{P}_{\bos} = \sqzb (\mbf{k}) $.
The total momentum operator is defined by
\begin{equation}
\mbf{P} = \overline{  \mbf{p}  \tens \one + \one \tens \mbf{P}_{\bos}  } , \label{tmo}
\end{equation}
where $\overline{X} $ denotes the closure of  operator $X$.
It is known that  $H^{0} (\kappa ) $ strongly commutes with ${P}_j $, $j=1,2,3$.
Then, it follows that for all $\mbf{a} \in \Rthree $,
\begin{equation} \label{TI}
  e^{i \mbf{a} \cdot \mbf{P}}   H^{0} (\kappa )   e^{-i \mbf{a} \cdot \mbf{P}}  = H^{0} (\kappa ) .
\end{equation}
In this sense, we say that $H^{0} (\kappa )$ has a translation invariant property. 
We assume  that  the potential decays as follows.
\begin{quote}
(A.2) (i) 
There exists $R_0 >0 $  such that for all $|x| \geq  R_0$,
\[
V (\mbf{x}) \leq - 4\frac{C_{\textrm{p}} + \delta_{\textrm{p}} }{|\mbf{x}|^2 } ,
\] 
where $ C_{\textrm{p}} =  \inf \left\{ \lang u , \Hp^0 u  \rang 
\left| \frac{}{} \right. u \in  C_{0}^{\, \infty} (\Rthree ) , \| u \| = 1, \textrm{supp} \, u \subset \left\{ \mbf{x} \in \Rthree \left| \right. 1 \leq  | \mbf{x}| \leq 2 \right\}  \right\} $ and 
  $\delta_{\textrm{p}} >0  $ is a constant. \\
(ii) It holds that
\[
\lim_{ | \mbf{x} | \to \infty } V (\mbf{x} ) = 0 . 
\]
\end{quote}
\begin{remark}
Note that the following condition (i)'   satisfies  the above condition  (i) : 
\begin{quote} 
(i)' There exist $R >0 $, $ C   >0 $ and $ 0< \mu <2 $ such that for all $|x| \geq  R$,
\[
V (\mbf{x}) \leq - \frac{C }{|\mbf{x}|^\mu } .
\] 
\end{quote}
\end{remark}

\noindent
An example of the potentials is the Coulomb potential   $V (x) = -\frac{\alpha}{|\mbf{x}|}$, $\alpha >0$.
Let $\chi, \overline{\chi } \in C^{ \, \infty } (\Rthree)$, which satisfy  
 (i) $\chi \geq 0$, $ \overline{\chi }  \geq 0$, (ii) $ \chi (x)^2  + \overline{\chi }  (x)^2 =1$, (iii)
$\chi (x) =1$ for $  |{x} | \leq 1$, and $ \overline{\chi }  (x) =1$ for  $ | {x} | \geq 2$.   Set $\chiR (x) = \chi  (\frac{x}{R})$ and $\overline{\chiR} (x) = \overline{\chi}  (\frac{x}{R})$.  Using $[ [X, Y ] , Y] = XY^2 - 2YXY + Y^2X$,  we have the IMS localization formula of $H (\kappa )$ (see \cite[Theorem 3.2]{CFKS08}):
\begin{align} \label{IMS}
H (\kappa ) 
=  \chiR H (\kappa ) \chiR +
\overline{\chi_R}  H (\kappa )  \overline{\chiR}
-\frac{1}{2M} \left( | \nabla \chiR |^2  + | \nabla \overline{\chiR} |^2 \right) .
\end{align}
\noindent
Let  
\begin{align*}
\Sigma_{{}_{ R}} (H (\kappa ) ) =  \inf_{
\Psi \in \ms{D} (H_0 ), \overline{ \chiR } \Psi \ne 0 } 
\frac{\lang  \overline{\chiR}   \Psi,  H (\kappa )  \overline{\chiR}   \Psi  \rang }{
\lang \overline{\chiR}   \Psi, \overline{\chiR}   \Psi \rang } ,
\end{align*}

\noindent
and 
\[
 \Sigma_{ \, \infty } (H (\kappa ) ) = \liminf\limits_{R \to \infty } \,\Sigma_{ R} (H (\kappa ) )  .
 \]
We set $ E(X) = \inf \sigma (X)$ for a self-adjoint operator $X$.    If $E (H ( \kappa ) ) <\Sigma_{ \, \infty } (H (\kappa ) ) $, we  say that  binding condition holds. The binding condition is first investigated in \cite{GLL01}, and it plays an important role in the analysis of the ground state. To show the binding condition, the ground state of $\Hp $ is usually used (e.g., \cite{GLL01, HS10}), but it needs some conditions (see Remark \ref{remark1}).   In this paper,  we use  the condition of the decay of the potential, not   the ground state, and prove the following. 

\begin{theorem} \textbf{(Binding condition)} \label{gapSigma}
Assume (A.1) and (A.2). Then,
\[
E(H(\kappa))  \; < \;  \Sigma_{ \, \infty } (H (\kappa ) ) .
\]
\end{theorem}

\noindent
We prove Theorem \ref{gapSigma} by the translation invariant property  and 
the decay of the potential.
By Theorem  \ref{gapSigma}, we can apply   \cite[Theorem 1]{Gr04} to $H(\kappa )$, and  the exponential decay around the lowest spectrum  follows; 
Let $\lambda \geq 0  $ and  $\beta > 0 $, which satisfy $E (H (\kappa )) < \lambda + \beta^2 < \Sigma_{ \, \infty } (H (\kappa ) )$. Then, it holds that
 \[
 \| e^{\beta |\mbf{x}| } E_{\lambda} (H (\kappa )) \| < \infty ,
 \]
 where $E_{\lambda} $ is the spectral projection of $H (\kappa ) $ onto $ [
 E (H (\kappa )), E (H (\kappa )) + \lambda ]$.   In particular, if $H (\kappa ) $ has the ground state $\Omega_{\kappa}$, then $ 
\| e^{\beta |\mbf{x}| } \Omega_{\kappa} \| \; < \infty $ (see also \cite[Proposition 3.17]{Hibook19}). 
For the exponential decay of the ground states, refer to \cite{GK23, Hi19} and reference therein.


\section{Proof of Theorem \ref{gapSigma}}
\begin{lemma} \label{elzero}
Let $u $ be a real-valued function in $C_{0}^{\, \infty} (\Rthree)$ and $\Psi \in \ms{D} (H_0 )$. 
Let $\Xi = u \Psi $. Then, 
\begin{align*} 
 \lang  \Xi , H (\kappa ) \Xi  \rang  = \lang  u \Psi  ,   (\Hp^0  u) \Psi \rang  + \lang \Xi  ,V \Xi \rang
+ \ell^{\, 0}_{\kappa }  ( u , \Psi  ) ,
\end{align*}
where $\ell^{\, 0}_{\kappa } ( u , \Psi  ) 
=\frac{1}{2M} \sum\limits_{j=1}^3 \lang   u ( \partial_{x_j}  \Psi ) ,   u ( \partial_{x_j}  \Psi) \rang
+ \lang u \Psi  ,u  ( \one \tens \Hb )  \Psi  \rang
    + \kappa \lang u \Psi  , u \HI \Psi   \rang   $.
\end{lemma}
\textbf{Proof.} 
We see that
\begin{equation}
\lang  \Xi , \Hp    \Xi \rang 
=\frac{1}{2M}  \sum_{j=1}^3 \lang  \partial_{x_j}  \Xi  ,  \partial_{x_j} \Xi  \rang 
+ \lang \Xi  ,V \Xi \rang .  \label{0905a0}
\end{equation}
Since $\Xi = u \Psi $, we have
\begin{align}
   \lang  \partial_{x_j}  \Xi  ,  \partial_{x_j} \Xi  \rang 
   & =  \lang   ( \partial_{x_j} u ) \Psi   ,  (  \partial_{x_j} u ) \Psi   \rang
 + \lang   ( \partial_{x_j} u ) \Psi   ,    u  ( \partial_{x_j} \Psi )   \rang \notag  \\
&  \qquad \qquad   +  \lang     u  ( \partial_{x_j} \Psi )  ,   ( \partial_{x_j} u ) \Psi    \rang
 +  \lang   u  (  \partial_{x_j}  \Psi )   ,    u   ( \partial_{x_j} \Psi  )  \rang . \label{0905a1}
\end{align}
For the third term of the right-hand side of (\ref{0905a1}),  we use  
the fact that $u$ is real-valued and integration by parts, and we have 
\begin{align}
   \lang      u  (  \partial_{x_j} \Psi) ,   ( \partial_{x_j} u ) \Psi   \rang
& =\lang      \partial_{x_j} \Psi , u  ( \partial_{x_j} u ) \Psi    \rang \notag \\
& = - \lang    \Psi ,   ( \partial_{x_j} u )^2  \Psi   \rang
- \lang   \Psi , u  ( \partial_{x_j}^2  u ) \Psi   , \rang 
- \lang    \Psi , u  ( \partial_{x_j} u )    ( \partial_{x_j} \Psi )   \rang  \notag \\
& =   - \lang    ( \partial_{x_j} u )   \Psi ,   ( \partial_{x_j} u )  \Psi   \rang
- \lang u  \Psi ,   ( \partial_{x_j}^2  u ) \Psi   , \rang 
- \lang   ( \partial_{x_j} u )   \Psi , u    ( \partial_{x_j} \Psi )   \rang .
\label{0905a2}
 \end{align}
Applying (\ref{0905a2}) to (\ref{0905a1}),  we have
\begin{equation}
\lang  \partial_{x_j}  \Xi  ,  \partial_{x_j} \Xi  \rang  = 
- \lang u  \Psi ,   ( \partial_{x_j}^2  u ) \Psi   , \rang  
+    \lang   u (   \partial_{x_j}  \Psi )  ,    u  ( \partial_{x_j} ) \Psi   \rang .  \label{0905a3}
\end{equation}
By (\ref{0905a0}) and (\ref{0905a3}) we have
\begin{align}
\lang  \Xi , \Hp  \Xi \rang = \lang u \Psi ,   (\Hp^0 u) \Psi \rang
+\frac{1}{2M} \sum_{j=1}^3 \lang   u  ( \partial_{x_j} \Psi ) ,   u ( \partial_{x_j} \Psi ) \rang 
 + \lang \Xi  ,V \Xi \rang, 
\end{align}
and the assertion follows. $\square $ \\

\noindent
For all $\mbf{y} \in \Rthree $, we set
\begin{equation}
U_{\mbf{y}} = e^{-i \mbf{y} \cdot \mbf{P}}  ,
\end{equation}
where $\mbf{P}$ is the total momentum operator defined in (\ref{tmo}).
Let $u \in C_{0}^{\, \infty} (\Rthree )$. 
Then,  it holds   that,  as a multiplication operator on $\ms{H}$,
\begin{equation}
U_{\mbf{y}} u (\mbf{x}) U^{\ast}_{\mbf{y}} =  u_{\mbf{y}} (\mbf{x}) ,  
\end{equation}
where   $  u_{\mbf{y}} (\mbf{x}) =  u( \mbf{x} - \mbf{y}) $. 

\begin{lemma} \label{Ey}
Let $ u \in C_{0}^{\, \infty} (\Rthree )$   with $\| u\| =1 $ and $\Psi \in \ms{D} (H_0 )$ with $\| \Psi \| =1$.
Let $\Xi_{\mbf{y}} = u  U_{\mbf{y}}^{\ast} \Psi$.
Then, 
\[
\int_{\Rthree}
\lang \Xi_{\mbf{y}} , H (\kappa ) \Xi_{\mbf{y}} \rang d \mbf{y}
\geq E (H(\kappa )) . 
\]
\end{lemma}
\textbf{Proof.}
We see that
\[
\int_{\Rthree}
\lang \Xi_{\mbf{y}} , H (\kappa ) \Xi_{\mbf{y}} \rang d \mbf{y}
\geq E (H(\kappa )) \int_{\Rthree}
\lang \Xi_{\mbf{y}} , \Xi_{\mbf{y}} \rang d \mbf{y} .
\]
Then, it follows that
\begin{align}
\lang \Xi_{\mbf{y}} , \Xi_{\mbf{y}} \rang
& = \lang  U_{\mbf{y}}  \Xi_{\mbf{y}} , U_{\mbf{y}}  \Xi_{\mbf{y}} \rang  \notag \\
& =  \lang  U_{\mbf{y}} u  U_{\mbf{y}}^{\ast} \Psi  , U_{\mbf{y}}  u  U_{\mbf{y}}^{\ast} \Psi   \rang \notag \\
& =\lang u_{\mbf{y}} (\mbf{x}) \Psi ,  u_{\mbf{y}} (\mbf{x})  \Psi   \rang   \notag \\
& = \int_{\Rthree} | u (\mbf{x} -\mbf{y})|^2 \lang \Psi (\mbf{x}) ,  \Psi (\mbf{x}) \rang_{\Fb} d \mbf{x}
\label{0906a1}.
\end{align}
From (\ref{0906a1}), we have 
\begin{align}
\int_{\Rthree}
\lang \Xi_{\mbf{y}} , \Xi_{\mbf{y}} \rang d \mbf{y} 
&=   \int_{\Rthree} \left\{ \int_{\Rthree} | u (\mbf{x} -\mbf{y})|^2 \lang \Psi (\mbf{x}) ,  \Psi (\mbf{x}) \rang_{\Fb} d \mbf{x} \right\} d \mbf{y}    \notag \\
&= \int_{\Rthree} \left\{ \int_{\Rthree} | u (\mbf{x} -\mbf{y})|^2  d \mbf{y}   \right\}  \lang \Psi (\mbf{x}) ,  \Psi (\mbf{x}) \rang_{\Fb} d \mbf{x}  \notag \\
& = \| u \|^2 \| \Psi \|^2  . \notag
\end{align}
Since $\| u \| =1$ and $\| \Psi \| =1 $,  the proof is obtained. $\square $ \\

\begin{lemma} \label{elHzero}
Let $u \in C_{0}^{\, \infty} (\Rthree)$ with $\| u \| =1$ and $\Psi \in \ms{D} (H_0 )$. 
Then, 
\[
\int_{\Rthree}
  \ell^{\, 0 }_{\kappa }    ( u , U_{\mbf{y}}^{\ast}  \Psi  )  d \mbf{y}
=   \lang \Psi , H^{0} (\kappa ) \Psi \rang .
\]
\end{lemma}
\textbf{Proof.}
Since $U_{\mbf{y}}$ strongly commutes with $\partial_{x_j} $, $j=1,2,3$, $ \one \tens \Hb$ and 
$\HI $, we see that 
\begin{align}
 \ell^{\, 0 }_{\kappa }  ( u , U_{\mbf{y}}^{\ast}  \Psi  )       =&  \frac{1}{2M} \sum\limits_{j=1}^3 \lang   u \partial_{x_j} U_{\mbf{y}}^{\ast}  \Psi  ,   u \partial_{x_j}
    U_{\mbf{y}}^{\ast}       \Psi \rang
+ \lang u U_{\mbf{y}}^{\ast} \Psi  , ( \one \tens \Hb )  u U_{\mbf{y}}^{\ast} \Psi   \rang \notag
 \\
&    \quad  + \kappa \lang u U_{\mbf{y}}^{\ast} \Psi  , \HI u U_{\mbf{y}}^{\ast}  \Psi  \rang  \notag \\
  &   = \frac{1}{2M} \sum\limits_{j=1}^3 \lang   U_{\mbf{y}}  u U_{\mbf{y}}^{\ast}  \partial_{x_j}  \Psi  ,  U_{\mbf{y}}  u  U_{\mbf{y}}^{\ast} \partial_{x_j}           \Psi \rang
+ \lang U_{\mbf{y}} u   U_{\mbf{y}}^{\ast} \Psi  , ( \one \tens \Hb ) U_{\mbf{y}} u U_{\mbf{y}}^{\ast} \Psi   \rang    \notag \\
&   \quad  + \kappa \lang U_{\mbf{y}} u U_{\mbf{y}}^{\ast} \Psi  , \HI U_{\mbf{y}} u U_{\mbf{y}}^{\ast}  \Psi  \rang \notag \\
&= \frac{1}{2M} \sum\limits_{j=1}^3 \int_{\Rthree} | u(\mbf{x}-\mbf{y})|^2 
  \lang  \partial_{x_j } \Psi (\mbf{x} ) ,  \partial_{x_j } \Psi (\mbf{x} )  \rang_{\Fb} d \mbf{x} \notag \\
&   \quad +  \int_{\Rthree} | u(\mbf{x}-\mbf{y})|^2 
\lang  \Psi (\mbf{x} ) , (\one \tens \Hb  + \kappa \HI )\Psi (\mbf{x} ) \rang_{\Fb }   d\mbf{x}  \notag .
\end{align}
Thus, we have
\begin{align*}
  \int_{\Rthree}
  \ell^{\, 0 }_{\kappa }  ( u , U_{\mbf{y}}^{\ast}  \Psi  )     d \mbf{y}  
  &  =  \int_{\Rthree} \left\{   \int_{\Rthree} |u(\mbf{x}-\mbf{y})|^2  d \mbf{y}  \right\} 
 \left( \frac{1}{2M} \sum\limits_{j=1}^3 \int_{\Rthree}   \lang  \partial_{x_j } \Psi (\mbf{x} ) ,  \partial_{x_j } \Psi (\mbf{x} )  \rang_{\Fb} \right) d \mbf{x}  \\
  &  \qquad  + \int_{\Rthree} \left\{   \int_{\Rthree} | u(\mbf{x}-\mbf{y})|^2  d \mbf{y}  \right\} \lang  \Psi (\mbf{x} ) , (\one \tens \Hb  + \kappa \HI )\Psi (\mbf{x} ) \rang_{\Fb }    d \mbf{x}  \\
  & = \| u \|^2 \left( \frac{1}{2M} \sum\limits_{j=1}^3 \int_{\Rthree}   \lang  \partial_{x_j } \Psi  ,  \partial_{x_j } \Psi  \rang   +\lang \Psi ,  (\one \tens \Hb  + \kappa \HI ) \Psi  \rang \right).  
\end{align*}
Since $ \frac{1}{2M} \sum\limits_{j=1}^3 \int_{\Rthree}   \lang  \partial_{x_j } \Psi  ,  \partial_{x_j } \Psi  \rang   
= \lang \Psi , \Hp^0 \Psi  \rang$, the proof is obtained. $\square $ \\

\noindent
The following lemma can be shown in the same way as  Lemma \ref{elHzero}, so the proof is omitted.

\begin{lemma} \label{HpV}
Let $ u \in C_{0}^{\, \infty} (\Rthree )$  and $\Psi \in \ms{D} (H_0 )$ with $\| \Psi \| =1$.
Let $\Psi_{\mbf{y}} =   U_{\mbf{y}}^{\ast} \Psi$. Then, it holds that
\begin{align*}
& \mathrm{(i)} \;   \int_{\Rthree} \lang u \Psi_{\mbf{y}} , (\Hp^0 u ) \Psi_{\mbf{y}} \rang d \mbf{y}
= \lang u, \Hp^0 u  \rang ,  \qquad \qquad \qquad   \qquad \qquad \qquad   \qquad \qquad \qquad \\
& \mathrm{(ii)} \;  \int_{\Rthree} \lang u \Psi_{\mbf{y}} , V u \Psi_{\mbf{y}} \rang d \mbf{y}
= \lang u,  V u  \rang  .
\end{align*}
\end{lemma}

\noindent 
From Lemma \ref{Ey},   Lemma \ref{elHzero} and Lemma\ref{HpV}, the next corollary follows.

\begin{corollary} \label{coro1}
Let $ u \in C_{0}^{\, \infty} (\Rthree )$   with $\| u\| =1 $ and $\Psi \in \ms{D} (H_0 )$ with $\| \Psi \| =1$.
Then, 
\[
E (H(\kappa)) \leq  \lang  u , \Hp^0 u \rang + \lang u , Vu \rang + \lang \Psi , H^0 (\kappa ) \Psi \rang . 
\]
\end{corollary}

\begin{remark} \label{remark1}
In Corollary \ref{coro1}, assume that $u$ is the ground state of $\Hp $, instead of  $u \in C_{0}^{\, \infty} (\Rthree ) $. Then it holds that
\[
E (H(\kappa)) \leq  E(\Hp ) \lang  u ,  u \rang  + \lang \Psi , H^0 (\kappa ) \Psi \rang .
\]
Note that in order to prove the above, we need the following additional conditions:  (i) $ u \Psi  \in L^2 (\Rthree)$,   (ii) $ (\partial_{x_j} u) \Psi  \in L^2 (\Rthree) $ and $   u   ( \partial_{x_j} \Psi  ) \in L^2 (\Rthree) $, $j=1,2,3$,  (iii) $(\Delta u) \Psi \in L^2 (\Rthree) $, (iv) $(Vu) \Psi \in L^2 (\Rthree) $ for  $\Psi \in \ms{D} (H_0) $. These conditions are needed in the proof of Lemma  \ref{elzero}.
\end{remark}

\begin{proposition}  \label{Binding}
Assume (A.1) and (A.2). Then, 
\[
E (H(\kappa )) \;  < \; E (H^0 (\kappa )) .
\]
\end{proposition}
\textbf{Proof.}
Let  $  u \in C_{0}^{\, \infty} (\Rthree )$. We assume that $\| u \| =1$ and supp$\, u \subset \left\{ \mbf{x} \in \Rthree \left| \right. 1 \leq  | \mbf{x}| \leq 2 \right\} $.
Let $ \uR (\mbf{x}) = \frac{1}{\sqrt{R}^3}u (\frac{\mbf{x}}{R})$, $R>0$. From the definition of $E(H^0 (\kappa ))$, it follows that 
for all $\epsilon >0 $, there exists $\Psi_{\epsilon } \in \ms{D} (H_0 )$ such that 
$\| \Psi_{\epsilon } \| =1$ and 
\[
\lang \Psi_{\epsilon }  , H^0 (\kappa) \Psi_{\epsilon } \rang <  E(H^0 (\kappa )) + \epsilon . 
\]
Then, by Corollary  \ref{coro1}, we have
\begin{align}
E (H(\kappa)) & \leq  \lang  \uR , \Hp^0 \uR \rang + \lang \uR , V \uR \rang + \lang \Psi_{\epsilon} , H^0 (\kappa ) \Psi_{\epsilon} \rang  \notag \\
&   < \lang  \uR , \Hp^0 \uR \rang + \lang \uR , V \uR \rang +   E(H^0 (\kappa )) + \epsilon . \label{0908a1}
\end{align}
We see that 
\begin{equation}
  \lang  \uR , \Hp^0 \uR \rang = \frac{c_{\mathrm{p}}}{R^2} , \label{0908a2}
\end{equation}
where $ c_{\mathrm{p}} =  \lang  u , \Hp^0 u \rang $. By (A.2) (i), for all $R \geq R_0$, it holds that
\begin{equation}
\lang \uR , V \uR \rang  \leq -4 \frac{C_{\textrm{p}} + \delta_{\textrm{p}} }{(2R)^2 } 
=  - \frac{C_{\textrm{p}} + \delta_{\textrm{p}} }{R^2 } . \label{0908a3}
\end{equation}
Applying (\ref{0908a2}) and (\ref{0908a3}) to (\ref{0908a1}), we have
\[
E (H(\kappa)) <   \frac{c_{\mathrm{p}}}{R^2}  - \frac{C_{\textrm{p}} + \delta_{\textrm{p}} }{R^2 }  + E (H^0 (\kappa ) ) + \epsilon    .
\]
From the definition of $C_{\textrm{p}}$,  we can choose  $u  \in C_{0}^{\, \infty} (\Rthree )$ such that  
$ c_{\textrm{p}} - ( C_{\textrm{p}} + \delta_{\textrm{p}} ) < 0$. 
Then we can take $\epsilon > 0 $ such that $  \frac{ c_{\textrm{p}} -(  C_{\textrm{p}} + \delta_{\textrm{p}} ) }{R^2 }   + \epsilon < 0 $. 
Thus, the proof is obtained. $\square$ \\

\begin{lemma} \label{liminfzero}
Assume (A.1) and (A.2). Then, 
\[
E (  H^0 (\kappa )) \; \leq  \;  \Sigma_{ \infty } (H (\kappa ) ) .
\]
\end{lemma}
\textbf{Proof.}
Let  $\epsilon >0 $.  From the definition of $ \Sigma_{ R} (H (\kappa ) )$, there exists $ \Psi_{\epsilon} \in \ms{D} (H_0 ) $ such that 
\begin{align}
  \frac{\lang \overline{\chiR}   \Psi_{\epsilon} ,  H (\kappa )  \overline{\chiR}   \Psi_{\epsilon}  \rang }{
 \lang \overline{\chiR}   \Psi_{\epsilon}  , \overline{\chiR}   \Psi_{\epsilon}  \rang } 
< \Sigma_{ R} (H (\kappa ) ) + \frac{\epsilon}{2}  . \label{0908b1}
\end{align}
Since we assume that $\lim\limits_{| \mbf{x} | \to \infty } V (\mbf{x}) =0 $ in (A.2) (ii), 
there exists $R(\epsilon ) > 0$ such that for all $|\mbf{x}| > R( \epsilon ) $, 
$|V (x)| < \frac{\epsilon}{2} $.  Then, we see that  for all $R > R (\epsilon )$,
\begin{align}
   \lang  \overline{\chiR}   \Psi_{\epsilon} ,  H (\kappa )  \overline{\chiR}   \Psi_{\epsilon}  \rang 
  & = \lang \overline{\chiR}   \Psi_{\epsilon} ,  H^0 (\kappa )  \overline{\chiR}   \Psi_{\epsilon}  \rang
+ \lang \overline{\chiR}   \Psi_{\epsilon} , V \overline{\chiR}   \Psi_{\epsilon}  \rang   \notag \\
& \geq E (H^0(\kappa )) \lang \overline{\chiR}   \Psi_{\epsilon}  , \overline{\chiR}   \Psi_{\epsilon}  \rang - \frac{\epsilon }{2}    \lang \overline{\chiR}   \Psi_{\epsilon}  , \overline{\chiR}   \Psi_{\epsilon}  \rang . \label{0908b2}
\end{align}
By (\ref{0908b1}) and (\ref{0908b2}),  we have
\[
E (H^0 (\kappa ))    \leq \Sigma_{ R} (H (\kappa ) ) + \epsilon. 
\]
From this we have
\[
E (H^0 (\kappa )) \leq \liminf_{R \to \infty } \,  \Sigma_{ R} (H (\kappa ) )  + \epsilon .
\]
Since $\epsilon > 0 $ is arbitrary,   the proof is obtained. $\square $ \\

$\;$ \\
\textbf{Proof of  Theorem \ref{gapSigma} }\\
It follows from Proposition \ref{Binding} and Lemma \ref{liminfzero}. $\square$

$\;$ \\
{\large {\textbf{Acknowledgements}}} \\
The author would like to thank  Professor  Fumio Hiroshima for his comments and advice.
 This work is supported by JSPS KAKENHI $20$K$03625$.  \\


\end{document}